\documentclass[12pt,aps,prb,preprint]{revtex4}  
\usepackage{amsmath}    
\usepackage{amsfonts}  
\usepackage{amssymb}
\usepackage{graphicx}   

\begin{document}

\title{Two forms of the action variable for the relativistic harmonic oscillator}

\author{M.K.Balasubramanya$^{a)}$}

\affiliation{Department of Physical and Environmental Sciences\\ Texas A\&M 
University-Corpus Christi\\6300 Ocean Drive \\ Corpus Christi, Texas, 78412 U.S.A.}   
\email{mirley.balasubramanya@tamucc.edu}

\date{\today}

\begin{abstract}

The frequency of a classical periodic system can be obtained using action variables 
without solving the dynamical equations. We demonstrate the construction of two 
equivalent forms of the action variable for a one dimensional relativistic harmonic 
oscillator and obtain its energy dependent frequency. This analysis of oscillation 
is compared with the traditional solution of the problem which requires the use of
hypergeometric series. 
\end{abstract}

\maketitle

\section{Key words} Hamilton-Jacobi theory, action variable, contour integration, relativity, 
simple harmonic oscillator, hypergeometric series.

\section{Introduction}               

Classical periodic systems can be analyzed elegantly in terms of their
action and angle variables which constitute a set of canonically 
conjugate momenta and coordinates. Action variables are proportional to 
$\oint p_{j}dq_{j}$, where $(q_{j},p_{j})$ are the system's coordinates 
and canonical momenta. For conservative systems they are constants of motion 
in the manner of angular momentum and energy. The frequencies of periodic systems 
can be found using the functional relationship between the action variables and 
total mechanical energy without requiring a complete solution of the dynamical 
equations. We first summarize here that theory and situate action variables within its matrix.
In Section II we apply this formalism to determine the frequency of a non-relativistic 
simple harmonic oscillator using two equivalent  contour integral definitions of the 
action variable. In Section III we extend this formalism to a relativistic simple
harmonic oscillator, obtain its frequency in two equivalent series representations and
demonstrate that it has the correct non-relativistic limit. In Section IV we derive
the expression for the period of this relativistic oscillator by direct integration and 
compare it with the one obtained using action variables.

The time evolution of a classical system is governed by its Hamiltonian $H$
which is a function of its coordinates $x_{i}$, the conjugate 
momenta $p_{i}$ and the time $t$. The dynamics of such a system is 
determined by Hamilton's equations of motion
\begin{eqnarray}
{\dot x}_{i}  =  \frac{\partial H(x_{i},p_{i},t)}{\partial 
p_{i}},\; \; \; \;  
{\dot p}_{i} =  -  \frac{\partial H(x_{i},p_{i},t)}{\partial x_{i}}. 
\label{hameq}
\end{eqnarray}     
In the case of a particle of mass $m$ moving in one dimension under 
the influence of a time independent potential energy function $V(x)$ the 
Hamiltonian  is given by $H =  \frac{p^{2}}{2m} + V(x)$. 
Such a Hamiltonian is a constant of the motion and is the total energy $E$ of 
the system. Thus,
\begin{eqnarray}
	\frac{p^{2}}{2m} + V(x)  =  E.
\label{orbit}
\end{eqnarray}

Canonical transformations are those transformations of one set of 
coordinate and momentum $(x,p)$ to another set $(X,P)$ that preserve 
the form of Hamilton's equations. One such transformation is generated by 
the generating function $W_{C}(x,P)$ (the suffix $C$ in this and similar variables
refers to the "classical" rather than the "quantum" nature of the mechanics
considered), whose arguments are the "old" coordinate, $x$, and the "new" momentum, $P$:
\begin{eqnarray}
 p  =  \frac{\partial W_{C}(x,P)}{\partial x},   \; \;
X = \frac{\partial W_{C}(x,P)}{\partial P}. &
\label{trans}
\end{eqnarray} 
If this transformation transforms the Hamiltonian into a function only of $P$, then, using ~(\ref{hameq}),
\begin{eqnarray}
\dot P  =  - \frac{\partial H(P)}{\partial X}  =  0 &  
\Rightarrow \; \; \; P(t)  =  P, \; {\rm a \; constant}, \nonumber&\\
\dot X  =  \frac{\partial H(P)}{\partial P}   =  V_{0},  \; \; {\rm a \; constant}, \; \; & 
\Rightarrow \; \; X(t)  =   V_{0}\: t + X_{0}. &
\label{transtwo}
\end{eqnarray}

Thus $X$ and $P$ evolve very simply in time. The former progresses 
linearly in time and the latter is a constant. 
$W_{C}(x,P)$, which generates a canonical transformation in which the 
transformed Hamiltonian is independent of the new coordinate $X$, is 
the Hamilton's characteristic function. It is related to Hamilton's 
principle function $S_{C}$ through $S_{C}(x,P) = W_{C}(x,P) - Et$ (which generates another
canonical transformation, not considered here) and, for 
the case of time independent Hamiltonians, satisfies the Hamilton-Jacobi 
equation obtained by using ~(\ref{trans}) in ~(\ref{orbit}):
\begin{eqnarray}
\frac{1}{2m}\left(\frac{\partial W_{C}(x,P)}{\partial x}\right)^2 + V(x) & = & E(P).
\label{hj}
\end{eqnarray}
 
The use of this method to solve the dynamical problem involves the 
following steps: (i) Define a suitable new constant momentum $P$,
(ii) Integrate Eq. ~(\ref{hj}) to obtain $W_{C}(x,E(P))$, 
(iii) Obtain $x(X,P)$ and $p(X,P)$ using Eq. ~(\ref{trans}), and
(iv) Express $X$ and $P$ in terms of the initial values $x_{0}, p_{0}$ and 
$t$. 

One particular form of Hamilton-Jacobi theory is particularly suited for 
the study of 
periodic motion. If an inspection of the Hamiltonian indicates that the 
motion 
is periodic, then by a particular choice of the new momentum $P$ we can 
evaluate the period of motion without obtaining a complete solution of the 
dynamical problem. The new canonically conjugate coordinate and momentum 
are chosen to be
$X = w, \; P = J_{C}$ with 
\begin{eqnarray}
	J_{C} & = & \frac{1}{2\pi}\oint p_{C}(x,E)dx,
\label{oldj}
\end{eqnarray}
where $p_{C}(x,E)$, from ~(\ref{orbit}), is $\sqrt{2m[E - V(x)]}$ 
and the integral in phase space is performed over one cycle of the 
periodic motion. $J_{C}$ is the classical action variable and $w$ the 
angle variable. Since $J_{C} = J_{C}(E)$ we can invert it to obtain $E = E(J_{C})$. 
From Eq.~(\ref{transtwo}) the time evolution of the new coordinate is 
$w(t) = {\omega}t + w_{0}$ where the constant "velocity" is
\begin{eqnarray}
\omega = \frac{\partial H(J_{c})}{\partial J_{C}} = \frac{\partial E(J_{c})}{\partial J_{C}}.
\label{wevolution}
\end{eqnarray}
It can be shown that $\omega$ is the angular frequency 
of this periodic motion \cite{goldstein}. Thus the mathematical problem of finding the 
frequency of classical periodic motion for a conservative system is 
reduced to that of performing the integral ~(\ref{oldj}), solving for $E$ 
to get $E(J_{C})$, and evaluating $\partial E/\partial J_{C}$. This is a 
simple and elegant method for evaluating the frequency of a system known to be periodic.
Charles-Eug\`{e}ne Delaunay (1816-1872) invented action and angle variables in the course of his study of periodicity 
of lunar motion \cite{Delaunay}. In the early days of quantum mechanics, Sommerfeld,
in his treatment of the motion of an electron in the hydrogen atom,  made use of the 
action variable, and evaluated it using a contour integral in the complex coordinate plane\cite{sommerfeld}. 
We follow his example and obtain the classical relativistic harmonic oscillator's frequency by evaluating 
the action variable using two different suitably defined contour integrals. There is an equivalent formalism in 
contemporary quantum mechanics where a quantum version of the action variable, also evaluated by contour integrals,
is employed to determine the energy eigenvalues of a bound quantum system. This has received 
attention in the last two decades \cite{LPR, Asiri, Bhalla}. 

An equivalent definition of $J_{C}$ is that it is the contour integral
\begin{eqnarray}
        J_{C} & = & \frac{1}{2\pi}\oint_{C} p_{C}(x,E) dx,
\label{compdef}
\end{eqnarray}
in the complex $x$ plane over a contour $C$ (specified next), where $p_{c}(x,E)$ is a complex valued function of the complex argument $x$, and is 
defined as a suitable branch of 
\begin{eqnarray}
	p_{c}(x,E) & = & \sqrt{2m\left[E - V(x)\right]}.
\label{pcdef}
\end{eqnarray}
The turning points $x_{1}$ and $x_{2}$, both real, are defined by 
$p_{c}(x_{1},E) = p_{c}(x_{2},E) = 0$. These are also
the branch points of $p_{c}(x,E)$ in the complex-$x$ plane. We choose a 
branch cut connecting $x_{1}$ and $x_{2}$ along the real axis. $p_{c}(x,E)$ 
is chosen as that branch of the square root which is positive along the bottom 
of the cut. The counterclockwise contour $C$ surrounds the two turning 
points $x_{1}$ and $x_{2}$ and the branch cut connecting them.  The integral in ~(\ref{compdef}) is performed by 
enlarging $C$ outward to another contour $\gamma$, expanding $p_{c}(x,E)$ in a Laurent 
series in an annulus that contains $\gamma$ and using Cauchy's residue theorem. 

An equivalent construction of the action variable has the 
$-\frac{1}{2\pi}\oint_{C'} x_{C}(p,E) dp$ form, where 
the clockwise contour $C'$ in the complex-$p$ plane encloses the two 
turning momenta, $p_{1}$ and $p_{2}$, and the branch cut connecting them 
along the real axis. This second form of the action variable is insufficiently explored in the literature, and there is no account
of the frequency determination of the relativistic oscillator using the action variable. 
We will demonstrate the use of this alternate form of the action variable for the harmonic oscillator in the
relativistic case, and for comparison and completeness, in the non-relativistic case also.

\section{Action variable for non-relativistic harmonic oscillator}

The Hamiltonian for the non-relativistic harmonic oscillator with spring 
constant $k$ is $H = \frac{p^{2}}{2m} + \frac{kx^{2}}{2}$. 
We will refer to the angular frequency of this oscillator,
$\sqrt{\frac{k}{m}}$, as ${\omega}_0$.
We first demonstrate the technique of constructing the classical action 
variable for this simple case using the two methods outlined earlier. We will 
extend it to the case of the relativistic harmonic oscillator in a very 
similar manner. 

\subsection{Action variable in the $\oint p dx$ form}
The two turning points of the oscillator, $x_{1}$ and $x_{2}$ for energy 
$E$, are obtained, using ~(\ref{orbit}), by setting 
$p_{C}(x_{1},E) = p_{C}(x_{2},E) = 0$. As functions of energy, they are
\begin{equation}
-x_{1} = x_{2} = \sqrt{\frac{2E}{k}}.
\label{classicaltp} 
\end{equation}
We write the momentum in the form 
\begin{equation}
p_{C}(x,E) \;=\; \sqrt{2m\:\left[E - \frac{kx^{2}}{2}\right]}\;=\; 
i\sqrt{mk} \; x \left[1 - \left(\frac{x_{2}}{x}\right)^2\right]^{\frac{1}{2}}
\label{pcpdxform}
\end{equation}
and extend it analytically in the complex-$x$ plane, 
with a branch cut connecting the turning points $x_{1}$ and $x_{2}$ along the real axis. 
The Laurent series for $p_{C}(x,E)$ in the region $|x| > x_{2}$ in powers 
of $\frac {x_{2}}{x}$ is obtained by using the binomial series for the square root:
\begin{equation}
p_{C}(x,E) = \sum_{j=1}^{\infty} a_{j} x^{3-2j} = i \sqrt{mk} \; x \left[1 - 
\frac{1}{2}\left(\frac{x_{2}}{x}\right)^2 - 
\frac{1}{8}\left(\frac{x_{2}}{x}\right)^4 \ldots\right]
\label{pcexpansionpdx}
\end{equation}
Deforming the contour $C$ outward into a circular contour $\gamma$ centered at the origin with radius greater than $x_2$, 
and evaluating the 
contour integral using Cauchy's residue theorem, we get
\begin{equation}
J_{C} = \frac{1}{2\pi}\:(2 \pi i)\:i \sqrt{mk}\:\left(a_{2} = 
-\frac{1}{2}{x_2}^2\right) = \frac{1}{{\omega}_{0}}E.
\label{Jinclasspdx}
\end{equation}
The angular frequency, from Eq.~(\ref{wevolution}), is
$\frac{\partial E}{\partial J_{C}}$, 
or $\omega = {\omega}_{0}$. For other kinds of oscillators where the 
relation $J_{C} = J_{C}(E)$ 
cannot be inverted to obtain $E = E(J_{C})$ in a closed form, we can use 
$\frac{1}{\omega} = \frac{\partial J_{C}}{\partial E}$. The non-dependence 
of $\omega$ on the simple harmonic oscillator's energy, and thus on its 
amplitude of oscillation,
arises from the purely linear relation between $E$ and $J_{C}$, 
which is characteristic of this simple classical periodic system, with its 
quadratic potential energy function. As we will see in the next section 
there is a richer structure to the relationship between $J_{C}$ and 
$E$ for a relativistic harmonic oscillator.

	\subsection{Action variable's $-\oint x dp$ form}

We demonstrate an alternate, but equivalent, form of the action variable 
for the simple harmonic oscillator. The turning momenta, $p_{1}$ and $p_{2}$, of the oscillator are 
defined, using ~(\ref{orbit}), by 
$x_{C}(p_{1}, E) = x_{C}(p_{2},E) = 0$, or  
\begin{equation}
-p_{1} = p_{2} = \sqrt{2 m E}.
\label{classturnmom}
\end{equation}
We write the coordinate in powers of $\frac{p_{2}}{p}$ as 
\begin{equation}
x_{C}(p,E) =  \sqrt{\frac{2}{k}}\sqrt{E - \frac{p^2}{2m}}	= 
\frac{-i}{\sqrt{mk}} p \left[1 - \left(\frac{p_{2}}{p}\right)^2\right]^{\frac{1}{2}}.
\label{classcoordfn}
\end{equation}
Using ~(\ref{classcoordfn}) we extend $x_{C}(p,E)$ into the complex-$p$ 
plane with a branch cut connecting $p_1$ and $p_2$ along the real axis. 
We choose that branch of $x_{C}(p,E)$ that is positive just above the 
branch cut. The alternate form of $J_{C}(E)$ is  
\begin{equation}
J_{C}  =  -\frac{1}{2\pi}\oint_{C'} x_{C}(p,E) dp,
\label{classaltJ}
\end{equation}
where the clockwise contour $C'$ encloses the branch cut. For $|p| > 
p_{2}$, we 
expand 
$x_{C}(p,E)$ in the Laurent series
\begin{equation}
x_{C}(p,E) = \sum_{j=1}^{\infty} a'_{j} p^{3-2j} = \frac{-i}{\sqrt{mk}} \; 
p \left[1 - \frac{1}{2}\left(\frac{p_{2}}{p}\right)^2 - \frac{1}{8} \left(\frac{p_{2}}{p}\right)^4 
\ldots\right]
\label{classxseries}
\end{equation}
Deforming the contour $C'$ outward into a circular contour ${\gamma}'$ centered at the origin with radius greater than $p_2$
and evaluating the integral in ~(\ref{classaltJ}), we get
\begin{equation}
J_{C} = -\frac{1}{2\pi}\:(2 \pi i)\:\frac{-i}{\sqrt{mk}} \:\left(a'_{2} = 
-\frac{1}{2}{p_2}^2\right) = 
\frac{1}{{\omega}_{0}}E
\label{classaltJeval}
\end{equation}
The expected result is that this $J_{C}(E)$ yields the angular frequency 
$\omega = \partial E/\partial J_{C} = {\omega}_{0}$.

\section{Relativistic harmonic oscillator}

The relativistic motion of the harmonic oscillator is governed by the 
Hamiltonian $H(x,p_{CR}) = \sqrt{{p_{CR}}^2 c^2 + m^2 c^4} + \frac{1}{2} k x^2$. 
The additional suffix $R$ indicates that the canonical momentum is relativistic. 
The total mechanical energy of the relativistic oscillator will be referred 
to as $E$, and $\tilde{E} = E - mc^2$ is its mechanical energy in excess of 
its rest mass energy. The dimensionless energy related parameter we will use 
is $\epsilon = \frac{\tilde{E}}{m c^2}$. The non-relativistic case is 
characterized by $\epsilon << 1$. 

	\subsection{$\oint p dx$ form of action variable}

The relativistic orbit equation, obtained from $H(x,p_{CR}) = \tilde{E} + mc^2$, is 
\begin{equation}
p_{CR}=\sqrt{2m\left[\tilde{E}-\frac{1}{2} k x^2\right]
\left[1+\frac{\tilde{E}-\frac{1}{2} k x^2}{2mc^2}\right]}.
\label{relorbit}
\end{equation}
There are four branch points, $x_{jR}, \; j=1,2,3,4$, in the complex-$x$ plane where 
$p_{CR}(x,\tilde{E})$ vanishes. Two of these are the physical turning 
points $x_{1R}$ and $x_{2R}$ given by 
\begin{equation}
-x_{1R} = x_{2R} = \sqrt{\frac{2\tilde{E}}{k}}.
\label{reltpx}
\end{equation} 
Their locations in the complex-$x$ plane are similar to those of the 
turning points $x_{1}$ and $x_{2}$ in the non-relativistic case, 
given by Eq.~(\ref{classicaltp}). The other two branch points of $p_{CR}(x,\tilde{E})$ 
are on the real axis at 
\begin{equation}
-x_{3R} = x_{4R} = \sqrt{\frac{2\tilde{E}}{k}} \sqrt{1 + 
\frac{2}{\epsilon}}.
\label{relunphystpx}
\end{equation} 
Their form indicates that these two branch points, unlike 
$x_{1R}$ and $x_{2R}$, are entirely 
relativistic in character. It is clear that $x_{2R}< x_{4R}$ for all energies. 
We choose a branch cut of $p_{CR}(x,\tilde{E})$ connecting $x_{1R}$ and 
$x_{2R}$ along the real axis. 
Two other branch cuts connect $x_{3R}$ and $x_{4R}$ with $x = \infty$ along the real axis. The 
branch of $p_{CR}(x,\tilde{E})$ that we choose for complex values of $x$ is positive just 
below the cut connecting $x_{1R}$ and $x_{2R}$. 

Using Eq.~(\ref{relorbit}) we rewrite $p_{CR}$ in the form
\begin{equation}
p_{CR}(x,\tilde{E})= \sum_{j=-\infty}^{\infty} A_{j} x^{3-2j} = i\sqrt{mk} 
\; \; x \left[1 - \left(\frac{x_{2R}}{x}\right)^2\right]^{\frac{1}{2}} 
\;(1+\epsilon)^{\frac{1}{2}}\left[1 - \left(\frac{x}{x_{4R}}\right)^2\right]^{\frac{1}{2}}.
\label{relmom}
\end{equation}
Comparing this with Eq. ~(\ref{pcpdxform}) we see that the multiplicative 
factor $ (1+\epsilon)^{\frac{1}{2}}[1 - (\frac{x}{x_{4R}})^2]^{\frac{1}{2}}$, 
which is very nearly 1 for low energies and for $|x| << |x_{4R}|$, 
modifies the non-relativistic $p_{C}(x,E)$ into the relativistic 
$p_{CR}(x,\tilde{E})$. 

The action variable $J_{CR}(E)$ is defined as  
\begin{equation}  
J_{CR}  =  \frac{1}{2\pi}\oint_{C_{R}} p_{CR}(x,E) dx,
\label{reljpdx}
\end{equation}
where the counterclockwise contour $C_{R}$ hugs the branch cut between $x_{1R}$ 
and $x_{2R}$. Expanding $p_{CR}$ in a Laurent series in the annulus $x_{2R} < |x| < x_{4R}$ we obtain 
\begin{eqnarray}
p_{CR}(x,E) =  i\sqrt{mk} \sqrt{1+\frac{\epsilon}{2}} \; \; x 
\left[
1 - \frac{{x_{2R}}^2}{2} 		
\left\{
1 - \frac{1}{8}\left(\frac{x_{2R}}{x_{4R}}\right)^2 
- \frac{1}{64} \left(\frac{x_{2R}}{x_{4R}} \right)^4 - \ldots  
\right\}
\frac{1}{x^2} 
\right] & \nonumber \\
+ {\rm \; powers \; of \; {\it x} \; other \; than} \; x^{-1}.
\label{laurentrelpdx}
\end{eqnarray}
The coefficient of $x^{-1}$ in this series, required 
for evaluating $J_{CR}$, is a series in powers 
of $\left(\frac{x_{2R}}{x_{4R}}\right)^2 = \frac{\epsilon}{2+\epsilon} < 1$, 
which is  $<< 1$ for low energies. Deforming 
the contour $C_{R}$ outward into the circular counterclockwise contour ${\gamma}_{R}$ centered 
at the origin with its radius less than $x_{4R}$,
and evaluating the integral in Eq. ~(\ref{reljpdx}), we get
\begin{equation}
J_{CR} = \frac{\tilde{E}}{{\omega}_{0}} \;
\sqrt{1+\frac{\epsilon}{2}}\; 
\left[1 - \frac{1}{8}\left(\frac{\epsilon}{2+\epsilon}\right) 
- \frac{1}{64} \left(\frac{\epsilon}{2+\epsilon}\right)^2 \ldots \right]
\label{reljexpr}
\end{equation}
Comparing $J_{CR}(\tilde{E})$ for $\epsilon << 1$ with $J_{C}(E)$ in Eq. 
~(\ref{Jinclasspdx}), we 
see that the relativistic action variable has the correct non-relativistic 
limit. 
For low energies, up to order $\epsilon$, we have $J_{CR} 
\approx 
\frac{\tilde{E}}{{\omega}_{0}}(1+\frac{\epsilon}{4})(1-\frac{\epsilon}{16}) 
\; 
\approx 
\frac{\tilde{E}}{{\omega}_{0}} \;(1 + \frac{3}{16} \epsilon)$. Finally, we 
calculate the relativistic angular frequency from the equation 
\begin{equation}
\frac{1}{{\omega}_{R}} = \frac{d J_{CR}}{d \tilde{E}} = 
\frac{1}{mc^2} \frac{d J_{CR}}{d \epsilon} = 
\frac{1}{{\omega}_{0}} \eta(\epsilon),
\label{relomegapdx}
\end{equation}
where the energy dependent relativistic factor $\eta(\epsilon)$ is
\begin{eqnarray}
\eta(\epsilon) = \frac{d}{d \epsilon} \; 
\left[\epsilon \;\sqrt{1+\frac{\epsilon}{2}}\; \left\{1 - 
\frac{1}{8}\left(\frac{\epsilon}{2+\epsilon}\right) 
- \frac{1}{64} \left(\frac{\epsilon}{2+\epsilon}\right)^2 \ldots 
\right\} \right].
\label{etafunction}
\end{eqnarray}
$\eta(\epsilon)$ is very nearly $1$ for low energies. This expression for 
${{\omega}_{R}}^{-1}$, valid for all energies, shows a decrease in frequency  
from the non-relativistic case due to time dilation, for an observer in the laboratory 
reference frame. We also see here the explicit dependence of this 
frequency on the oscillator's energy, and therefore, on its amplitude.
This expression is equivalent to the series representation of the period of this relativistic oscillator 
obtained by direct integration and shown in Section IV (See Eq. 
~(\ref{integratedtau})). 
Up to order $\epsilon$, it reduces to 
$\frac{1}{{\omega}_{R}} \approx \frac{1}{{\omega}_{0}} (1 + \frac{3}{8} 
\epsilon) $. 

	\subsection{$-\oint x dp$ form of action variable}

Starting with Eq. ~(\ref{relorbit}) and extending it into the complex $p$ 
plane we write the coordinate as
\begin{equation}
x_{CR} = -\sqrt{\frac{2}{k}}\sqrt{E - \sqrt{p^2c^2 + m^2c^4}}  .
\label{relxintermsofp}
\end{equation}
We see that there are two turning momenta $p_{jR}, \; j=1,2$,  given by
$E - \sqrt{p_{jR}^2c^2 + m^2c^4} = 0$, or
\begin{equation}
-p_{1R} = p_{2R} = \sqrt{2m\tilde{E}\left(1+\frac{\epsilon}{2}\right)}.
\label{turnmomrel}
\end{equation}
These two turning momenta, with a dependence on the relativistic factor 
$\sqrt{1+\frac{\epsilon}{2}}$ invite comparison with their 
non-relativistic counterparts in Eq.~(\ref{classturnmom}). They are also 
branch points of $x_{CR}(p,\tilde{E})$ and we choose one of its branch 
cuts from $p_{1R}$ to $p_{2R}$.  The presence of 
$\sqrt{p^2 c^2 + m^2 c^4}$ in Eq.~(\ref{relclascoordfn}) produces two 
additional branch points of $x_{CR}(p,\tilde{E)}$ of relativistic origin, 
given by 
$-p_{3R} = p_{4R} = i m c$. We choose the second set of branch cuts along 
the imaginary axis connecting each of $p_{3R}$ and $p_{4R}$ to $p = \infty$. 
Further, we choose the branch of $x_{CR}(p,\tilde{E})$ that is positive just above 
the branch cut connecting $p_{1R}$ and $p_{2R}$.

The coordinate in Eq.~(\ref{relxintermsofp}) can be rewritten, using the 
turning momenta, in a form similar to the non-relativitic coordinate Eq.~(\ref{classcoordfn}), and suitable for Laurent expansion, as
\begin{equation}
x_{CR}(p,\tilde{E}) = \sum_{j=-\infty}^{\infty} A'_{j} p^{3-2j} = 
\frac{-i}{\left[mk(1 + \frac{\epsilon}{2})\right]^{\frac{1}{2}}} \;p 
\sqrt{1 - \left(\frac{p_{2R}}{p}\right)^2} \;f(p^2,\epsilon), 
\label{relclascoordfn}
\end{equation}
where
\begin{equation}
f(p^2,\epsilon) = \frac	{\left[(1+\epsilon) - \left\{1 - \left(\frac{p}{{p_{4R}}}\right)^2    
\right\}^{\frac{1}{2}}\right]^{\frac{1}{2}} }	
{\left[-\epsilon  \left(\frac{p}{p_{2R}}\right)^2 \left\{1 -  
\left(\frac{p}{p_{2R}}\right)^2\right\}\right]^{\frac{1}{2}}}.
\label{fofpsquare}
\end{equation}
$f(p^2,\epsilon)$ is analytic for $|p^2| < |{p_{4R}}^2|$, is $1$ at $p^2=0$, and thus its Laurent series 
consists only of the non-negative powers of $p^2$.
We notice that, in this form, $x_{CR}(p,\tilde{E})$, apart from 
multiplicative constants, is but a product of $p$ and two varieties of
binomial series, one in powers of $\frac{p_{2R}}{p}$ and the other in 
$\frac{p}{p_{4R}}$, which converge uniformly in the annulus $p_{2R} < p < 
p_{4R}$. Expanding the square roots in Eq.~(\ref{relclascoordfn}) we get 
\begin{equation}
x_{CR} = -\sqrt{\frac{-2\tilde{E}}{k}} \; \frac{p}{p_{2R}} \;
\left[ 
1 - \frac{1}{2} \left( \frac{p_{2R}}{p} \right)^2 - \frac{1}{8}  \left( 
\frac{p_{2R}}{p} \right)^4  - \frac{1}{16}  \left( \frac{p_{2R}}{p} \right)^6 
\ldots  
\right]\left[1 + \sum_{j=1}^{\infty} f_{j}(\epsilon) p^{2j}\right],
\label{relcoordbinomialseries}
\end{equation}
where $f_{j}(\epsilon)$, which are inversely proportional to $p_{4R}^{2j}$, are the expansion coefficients in the Laurent series 
of $f(p^2,\epsilon)$. 
The coefficient of $\frac{1}{p}$ necessary for evaluating the 
residue is a power series in the parameter
$\left(\frac{p_{2R}}{p_{4R}}\right)^2 = 2\epsilon \left(1+\frac{\epsilon}{2}\right)$, which, for
low energies, is of order $\epsilon$:

The alternate definition of $J_{CR}(\tilde{E})$ is  
\begin{equation}
J_{CR}  =  -\frac{1}{2\pi}\oint_{C'_{R}} x_{CR}(p,\tilde{E}) dp,
\label{jcrxdp}
\end{equation}
where the clockwise contour $C'_{R}$ embraces the branch cut connecting $p_{1R}$ and 
$p_{2R}$. We expand this contour outward into the circular clockwise contour ${\gamma}'_{R}$ 
centered at the origin with a radius less than $p_{4R}$.
Evaluating the integral in Eq.~(\ref{jcrxdp}) on 
${\gamma}'_{R}$ using Eq.~(\ref{relclascoordfn}) we get
\begin{equation}
J_{CR}(\tilde{E}) = \frac{\tilde{E}}{{\omega}_{0}} \;\sqrt{1 + 
\frac{\epsilon}{2}} \; \left[1 - \frac{1}{16}\epsilon + \frac{7}{256} 
{\epsilon}^2 + \frac{1}{128} {\epsilon}^3 \ldots \right]
\label{jcrxdpevaluated}
\end{equation}

This expression for the relativistic action variable is a different 
series representation than the 
one in Eq. ~(\ref{reljexpr}), and reduces to the non-relativistic $J_{C}$ 
in Eq.~(\ref{classaltJ}) for $\epsilon << 1$. The 
angular frequency, ${\omega}_{R}$, is given by 	
\begin{equation}
\frac{1}{{\omega}_{R}} \;=\; \frac{d J_{CR}}{d \tilde{E}} \;= \;
\frac{1}{mc^2} \frac{d J_{CR}}{d \epsilon} \; = \;
\frac{1}{{\omega}_{0}} \frac{d}{d \epsilon} \left[\epsilon \:
\sqrt{1 + \frac{\epsilon}{2}} \; 
\left\{1 - \frac{1}{16}\epsilon + \frac{7}{256} 
{\epsilon}^2 + \frac{1}{128} {\epsilon}^3 \ldots \right\} \right]
\label{relomegaxdp}
\end{equation}
Truncating this series to the first order in $\epsilon$ for low energies, we recover the 
previous result, 
$\frac{1}{{\omega}_{R}} \approx \frac{1}{{\omega}_{0}} (1 + 
\frac{3}{8} \epsilon)$.

\section{Period of relativistic harmonic oscillator - Traditional treatment}

We evaluate here the period of the relativistic oscillator by direct 
integration. The period $\tau$ of a relativistic harmonic oscillator, for 
all energies, 
can be obtained in a closed form by integrating $\oint \frac{dx}{\dot{x}}$ 
over one cycle 
of the motion. For a potential energy function $V(x)$ which has the form 
of a symmetric well and is even in $x$,
\begin{equation}
\tau = \frac{4}{c} \int_{0}^{x_{2R}} {\frac{[E - V(x)] dx}{\sqrt{\left[E - 
V(x)\right]^2 - m^2 c^4}}}
\label{directperiod}
\end{equation}
$x_{2R}$ is the relativistic turning point on the right given by
$E - m c^2 - V(x_{2R}) = 0$. For the harmonic oscillator, with $V(x) = 
\frac{1}{2} k x^2$, the period integrates to 
\begin{equation}
\tau \;=\; \frac{2\pi}{c} \sqrt{\frac{2}{k}} \left[\sqrt{\tilde{E} + 2 m 
c^2} 
\left(1 - 
\frac{1}{4}{\kappa}^2  -  \frac{3}{64}{\kappa}^4 \ldots \right)\; \; \; - 
\; \; \; \frac{m c^2}{\sqrt{\tilde{E}+2 m c^2}} 
F_{1}^2\left(\frac{1}{2},\frac{1}{2} \; |1| \;
{\kappa}^2\right) \right],
\label{integratedtau}
\end{equation}
with $\kappa = \sqrt{\frac{\tilde{E}}{\tilde{E}+2 m c^2}}$, and the 
hypergeometric 
series $F_{1}^2$ given by
\begin{eqnarray}
F_{1}^2(a,b \;|c|\;z) = 1 + \frac{ab}{c} z + \frac{a(a+1) b(b+1)}{c(c+1)}
z^2 \ldots
\nonumber
\end{eqnarray}
 
For the weak relativistic case, where $\tilde{E} << mc^2$, we retain terms 
up to order ${\kappa}^2$, and obtain
\begin{equation}
\tau \approx 2\pi \sqrt{\frac{m}{k}} \; \left[ 
2(1+\frac{\epsilon}{2})^{\frac{1}{2}}
\left(1 - \frac{1}{4}{\kappa}^2\right) - \left(1+\frac{\epsilon}{2}\right)^{-\frac{1}{2}} 
\left(1 + \frac{1}{4}{\kappa}^2\right) \right]
\label{relativisticperiod}
\end{equation}
Further, ${\kappa}^2 \approx \frac{\epsilon}{2} (1 - 
\frac{\epsilon}{2})$ and 
\begin{equation}
\tau \approx 2\pi \sqrt{\frac{m}{k}} \; \left[1 + \frac{3}{8} \epsilon \right] \; \; 
\Rightarrow \omega \approx {\omega}_{0} \left[ 1 - \frac{3}{8} \epsilon \right]
\label{weakrelativisticperiod}
\end{equation}

\section{Conclusion}

We have shown the utility of the contour integral definition of the action variable in determining the
frequency of the relativistic harmonic oscillator. The formalism is easily extended to periodic systems in two and three
dimensions for separable systems. The non relativistic frequency emerges naturally for the $\epsilon << 1$ case. 
A series representation of the frequency of any other relativistic periodic system can be similarly obtained. 
Further, other Hamiltonian models of perodic systems lend themselves to this analysis. The central problem in this development is 
the identification of branch points of the temporally varying quantity (e.g., $p(x,E)$ or $x(p,E)$) expressed as a function of its 
conjugate variable and other constants of motion. 
There are four such points each in the two versions of the 
relativistic oscillator considered here, with the expansion parameter for the frequency being the ratio of a "near" and a "far" branch point. 
The case of a general periodic system in one dimension is characterized by its frequency depending on several such ratios of magnitude 
less than 1, with each ratio depending on the system's energy.

\end{document}